\documentclass[preprint,amsmath,amssymb]{revtex4}
\usepackage{pgfplots}
\usepackage{booktabs}
\usepackage{graphicx}
\usepackage{dcolumn}
\usepackage{bm}
\usepackage[skip=2pt,font=footnotesize]{caption}
\usepackage{amsmath}
\usepackage{mathrsfs}
\usepackage[english]{babel}
\usepackage{float}

\begin{document}
\normalsize
\title{Quasi-local stress-tensor formalism and the Casimir effect\\}
\author{ Borzoo Nazari  \footnote{borzoo.nazari@ut.ac.ir}}
\affiliation{University of Tehran, Tehran, Iran, P.O.Box: 11155-4563}

\begin{abstract}
We apply the quasi-local stress-energy tensor formalism to the Casimir effect of a scalar field confined between conducting planes located in a static spacetime. We show that the surface energy vanishes for both Neumann and Dirichlet boundary conditions and consequently the volume Casimir energy reduces to the famous zero point energy of the quantum field, i.e. $E^{vol.}=\sum\frac{\hbar \omega}{2}$. This enables us to reinforce previous results in the literature and extend the calculations to the case of massive and arbitrarily coupled scalar field. We found that there exists a first order perturbation correction to the Casimir energy contrary to previous claims which state that it vanishes. This shows many orders of magnitude greater than previous estimations for the energy corrections and makes it detectable by near future experiments.
\end{abstract}
\maketitle
\section{Introduction}
The Casimir effect has been the subject of numerous studies since its discovery in 1948 by H.B.G. Casimir \cite{Casimir}. The effect appears as an attractive/repulsive force on boundaries in the presence of some quantum field. The periodic or anti-periodic boundary condition may arise due to the presence of a material boundary or when the spacetime admits some compact dimensions without necessarily having a material boundary \cite{DeWitt,Ford}.
Thus, in the context of extra dimensions, many studies concerning the role of Casimir force in the vacuum structure of the underlying theories have been accomplished. See \cite{Teo2} and references therein. Some other studies refer to the long standing question of the effect of the gravitational field on the quantum vacuum energy, i.e. the zero point energy of quantum field theory. The Casimir effect in curved spacetime has been analysed by many authors \cite{Bezerra,SorgeNew,Muniz1,Muniz2,Geyer,Bimonte,Bimonte1,Sorge,Milton,Esposito,Napolitano,NouriNazari,Dowker} and unfortunately most of the calculations have been ended up with implicit and complicated formulas for the stress energy tensor including the energy and pressure respectively. A brief literature review on explicit results would be instructive.\\
Sorge \cite{Sorge} calculated corrections to the Casimir energy of a massless minimally coupled scalar field between two parallel planes in Schwarzschild spacetime using an expanded isotropic metric of the form $g_{\mu\nu}=diag(1+2\gamma+2\lambda z,-1-2\gamma-2\lambda z,-1-2\gamma-2\lambda z,-1-2\gamma-2\lambda z)$ and found the Casimir energy per unit volume as
\begin{eqnarray}\label{eq1}
E=-\frac{\pi^2}{1440 l_p^3}(1-\lambda l_p),\;\;\; \hbar=c=1,
\end{eqnarray}
in which $l_P$ is the proper distance between the planes and $\gamma=O(\epsilon),\; \lambda=O(\epsilon^2)$ are small parameters. In a recent paper \cite{Sorge2019}, the same author confirmed this result through different approach of the Schwinger's action principle. Note that the energy does not depend on $\gamma$.
\\
Bimonte et al \cite{Bimonte}, see specially the three published errata, calculated the electromagnetic energy-momentum tensor for parallel planes in Fermi coordinates which is described by $g_{\mu\nu}=diag(1+2 g z,-1,-1,-1)$ for some small parameter $g$ and concluded that the total Casimir energy decreases according to
\begin{eqnarray}\label{eq2}
E=-\frac{\pi^2}{720 l^3}(1+\frac{l}{2}g),\;\;\; \hbar=c=1.
\end{eqnarray}
Fermi coordinate, which describes the spacetime of a hovering observer in a static spacetime, is equivalent to the well-known accelerated observer of flat spacetime via weak principle of equivalence. In Ref.\cite{Bimonte1} they confirmed the above result and extend the calculations to next order of perturbations using such equivalency. Using the same method, Esposito et al \cite{Esposito} and Napolitano et al \cite{Napolitano} repeated the calculations for the case of a massless scalar field under Dirichlet and Neumann boundary conditions (D.N. B.C.s) on the planes. The obtained correction, as expected, was the same as equation (\ref{eq2}) times a factor of $\frac{1}{2}$ due to the fact that the electromagnetic field has two degrees of freedom relative to that of the scalar field.
Nazari \cite{BorzooEPJC} extended the calculations for scalar and electromagnetic fields to the more general metric of the form \cite{BorzooEPJC}
\begin{eqnarray}\label{eq3}
ds^2 = (1 +2\gamma_0 +2 \lambda_0 z) dt^2- (1+2\gamma_1 +2 \lambda_1 z)\left(dx^2+dy^2+dz^2\right),
\end{eqnarray}
in which $\lambda_0 z,\lambda_1 z,\gamma_0$ and $\gamma_1$ were small. It was shown in \cite{BorzooEPJC} that the metric (\ref{eq3}) describes any static spacetime (after expanding in the space between the planes) and the total energy for D.N. B.C.s was found.

Another trend of computations in the literature concerns the force exerted to the apparatus by the gravitational field in which the planes are located in \cite{Calloni}. In a series of papers Fulling et al \cite{Fulling}, Milton et al \cite{Milton} and Shajesh et al \cite{Shajesh} considered the force to parallel planes seeking confirmation of the weak principle of equivalence in Fermi coordinates. They showed the Casimir energy, and probably the quantum vacuum, gravitates just as required by the principle of equivalence and confirmed the energy (\ref{eq2}). However, they failed to derive the correct formula for the force. Bimonte \cite{Bimonte2} found a way according to which one can find the correct formula for the force exerted to any matter configuration by a constant gravitational field.

Calculations related to the Casimir energy in curved spacetime are complex and lengthy.
As has been pointed out in a recent work \cite{BorzooCQG}, the usual procedure consists of calculating a typical summation
\begin{eqnarray}\label{eq4}
E=\frac{\hbar}{2} \int \sum_{\omega_n,k} <T_{00}(\omega_n,k,\textbf{x})> dV
\end{eqnarray}
Many authors \cite{Bezerra,SorgeNew,Muniz1,Muniz2,Sorge,Bimonte,Bimonte1,Esposito,Napolitano,NouriNazari,Sorge2019,Lima,Blasone,Buoninfante,Lambiase} performed such a procedure to find the Casimir energy for D.N. BCs in various spacetimes. The point is that none of the above studies distinguished between the volume Casimir energy and the total one which is composed of a surface part resides on boundaries and a volume part in the bulk. Employing the quasi-local approach to the gravitational action,
Saharian \cite{Saharian2} had previously shown such a point. Accordingly, we know that the volume energy always satisfies
\begin{eqnarray}\label{eq5}
E^{vol.}=\frac{\hbar}{2} \int \sum_{\omega_n,k} \omega_n d^2k- E^{surf.},
\end{eqnarray}
in any static spacetime. In addition, this is true regardless of the shape of the boundary. Thus, most of the complexities encountered in calculation of equation (\ref{eq4}) is due to the surface term $E^{surf.}$ because we know that the first term in the right side of (\ref{eq5}) is nothing but the zero point energy of the corresponding quantum field which can be computed much easier.

Our purpose is to apply the quasi-local formalism to the Casimir effect for parallel plates. This approach has many advantages. First of all, the Casimir energy is a phenomenon under the presence of boundaries. Thus, it is natural to investigate it using the formalism of gravitational action on manifolds with boundaries. Second, it is simple and unambiguous and notably reduces the calculations in a way that enables us to extend previous results found by other studies
to the case of massive and arbitrarily coupled scalar field. Moreover, we find that the correction to the Casimir energy do not vanish within first order perturbation calculations, contrary to previous studies in the literature. Therefore, it could be measured by precise experiments in the near future.

The structure of the paper is as follows. Most of section II is a review of the Brown-York notation and formalism for stress tensor on manifolds with boundaries.
In section III we show that the surface energy vanishes for N.D. BCs. An important discussion and calculation in justification of the appearance of the first order corrections into the Casimir energy is presented. Conclusion is the final section.

\section{The stress tensor for manifolds with boundaries}
Suppose a compact $(D+1)-$dimensional spacetime manifold $M$ along with a metric $g_{\mu\nu}$ and boundary $\partial M$. The spacetime has been foliated by typical spacelike hypersurfaces $\Sigma$ each of them has a boundary $\partial \Sigma$. As Fig.1 shows, $\partial M=\partial M_s\cup \Sigma_1\cup \Sigma_2$ where $\Sigma_1$ and $\Sigma_2$ are $D-$dimensional initial and final spacelike hypersurfaces. Evidently, $\partial M_s$ is nothing but the evolution of $\partial \Sigma$, i.e. $\partial \Sigma=\Sigma \cap \partial M_s$. Normal vector to $\partial M$ is denoted by $n^\mu$ and $n^\mu n_\mu=\epsilon$ in which $\epsilon=1$ for spacelike hypersurfaces $\Sigma_1,\; \Sigma_2$ and $\epsilon=-1$ for $\partial M_s$. Unit timelike vector $u^\mu$ is normal to the hypersurfaces $\Sigma$. Thus, $u^\mu=n^\mu$ on $\Sigma_1$ and $u^\mu=-n^\mu$ for $\Sigma_2$. The unit vector $\overline{n}^\mu$ is defined as the vector field normal to $\partial \Sigma$ and lies on $\Sigma$. $\overline{u}^\mu$ is also normal to $\partial \Sigma$ and tangent to $\partial M_s$. For now we assume $M$ is not orthogonally foliated.
\begin{figure}
\begin{minipage}[c]{0.8\linewidth}
 \includegraphics[width=0.5\linewidth]{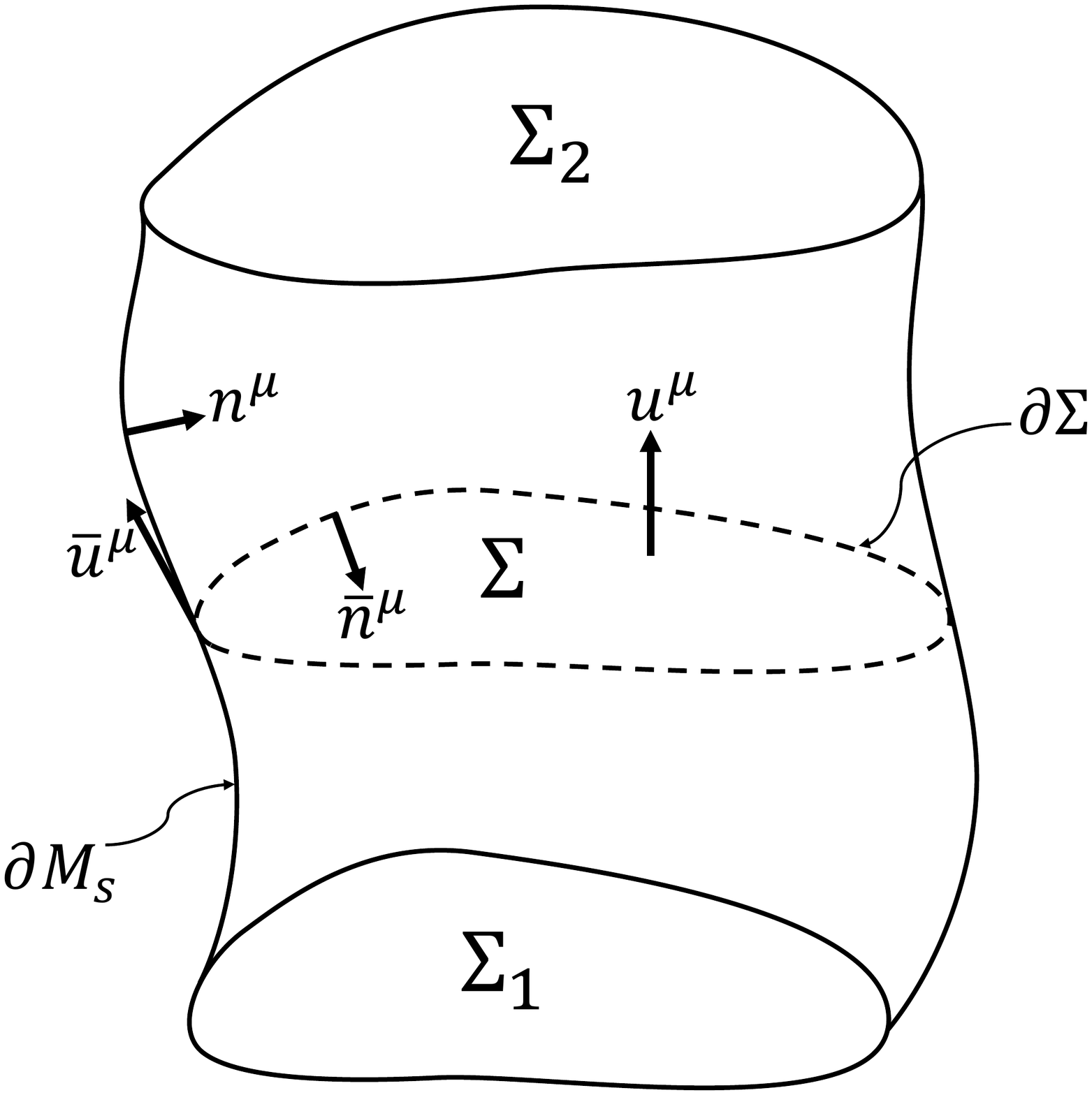}\label{fig:apparatus}
 \caption{Schematics of spacetime compact manifold foliated with hypersurfaces $\Sigma$. The boundary is defined by $\partial M=\partial M_s\cup \Sigma_1\cup \Sigma_2$. Unit vectors $n^\mu$ and $u^\mu$ are normal to $\partial M$ and $\Sigma$. Two vectors $\overline{n}^\mu$ and $\overline{u}^\mu$ are both normals to $\partial \Sigma$, the former tangent to $\Sigma$ and the later tangent to $\partial M_s$. }
 \end{minipage}
\end{figure}
The action describing a massive arbitrarily coupled scalar field consists of bulk and surface parts which are given by \cite{Saharian2}
\begin{subequations}
\begin{align}
&S=S_b+S_s, \label{eq6a}\\
&S_b=\frac{1}{2}\int _{M}d^{D+1}x\sqrt{|g|}\left( \nabla _\mu \phi \nabla ^\mu\phi -m^2 \phi ^2- \zeta R \phi ^2\right) ,  \label{eq6b} \\
&S_s=-\epsilon \int _{\partial M}d^Dx\sqrt{|h|}\left( \zeta \phi^2K +m_s \phi ^2 \right). \label{eq6c}
\end{align}
\end{subequations}

In the surface action, $h$ is the determinant of the projection tensor $h_{ik}=g_{ik}-\epsilon n_i n_k$ and $K$ the scalar obtained by contraction of extrinsic curvature $K_{ik}=h^l_ih^m_k\nabla _ln_m$. The parameter $m_s$ is free to keep the consistency of boundary conditions. The first term in the surface action has a counterpart in general relativity known as the Gibbons-Hawking term and its presence is necessary to obtain correct equations of motion when the induced metric held fix on the boundaries. Another point which is worth to mention is related to the case when the unit normal $n^\mu$ abruptly changes direction on the surface, i.e. when the surface is non-smooth on some edges. In this case, there must be taken into account some extra terms in the action to get rid of such potentially divergent terms. In Fig.1 sharp edges are the boundaries of $\Sigma_1$ and $\Sigma_2$. Therefore, we assume there is no such non-smoothness as we are interested here to take variation when the metric and field are held fixed on $\Sigma_1$ and $\Sigma_2$.

\subsection{Variation with respect to metric}
Taking variation of (\ref{eq6a}) with respect to metric results in \cite{Saharian2}
\begin{eqnarray}\label{eq7}
\delta _{g}S &=&\frac{1}{2}\int_{M}d^{D+1}x\sqrt{|g|}\, \delta g^{\mu\nu}T_{\mu\nu}^{\mathrm{(vol)}}+
\frac{1}{2}\int_{\partial M_s}d^{D}x\sqrt{|h|}\, \delta g^{\mu\nu}\tau _{\mu\nu} \nonumber \\
&&+\frac{1}{2}\int_{\partial M_s}d^{D}x\sqrt{|h|} \zeta D_{\lambda}(\phi ^{2}n^{\mu}h^{\lambda\theta}\delta g_{\mu\theta})
\end{eqnarray}
in which the volume and surface energy momentum tensors $T_{\mu\nu}^{(\mathrm{vol})}$ and $T_{\mu\nu}^{\mathrm{(surf)}}=\tau _{\mu\nu}\delta (x;\partial M_s)$ are defined by
\begin{subequations}
\begin{align}
&T_{\mu\nu}^{\mathrm{(vol)}}=\nabla _{\mu}\phi \nabla _{\nu}\phi -\frac{1}{2}%
g_{\mu\nu}g^{\lambda\theta}\nabla _{\lambda}\phi \nabla _{\theta}\phi
+\frac{m^2}{2}\phi ^2 g_{\mu\nu}  \nonumber \\
&\;\;\;\;\;\;\;\;\;\;\;\;\;\;\;\;\;\;\;\;\;-\zeta \phi ^{2}G_{\mu\nu}+\zeta g_{\mu\nu}g^{\lambda\theta}\nabla _{\lambda}\nabla _{\theta}\phi ^{2}-\zeta \nabla
_{\mu}\nabla _{\nu}\phi ^{2}  \label{eq8a}\\
&T_{\mu\nu}^{\mathrm{(surf)}}= \left\{ \zeta \phi ^{2} K_{\mu\nu}-h_{\mu\nu}\left( \zeta \phi^{2}K + \zeta n^{\theta}\nabla _{\theta}\phi^{2}+ m_s \phi ^2\right) \right\}  \delta (x;\partial M_s).
\label{eq8b}
\end{align}
\end{subequations}
One can easily check that $T_{\mu\nu}^{(1)}$ is divergence-free onshell, i.e. when the Klein-Gordon equation of motion holds. An expectable feature of surface energy momentum tensor is that it satisfies $T_{\mu\nu}^{\mathrm{(surf)}} n^\mu=0$.

On ground of discussion after (\ref{eq6c}), the last term in (\ref{eq7}) is nothing but the divergent term representing possible non-smoothness due to edges on boundaries. We drop it by now.

\subsection{Variation with respect to $\phi(x)$}
After taking variation with respect to $\phi(x)$, the equations of motion are found to be
\begin{subequations}
\begin{align}
&g^{\mu\nu}\nabla _{\mu}\nabla _{\nu}\phi +m^2 \phi +\zeta R\phi
=0, \label{eq15a}\\
&2(\zeta K +m_s)\phi + n^\mu\nabla _\mu \phi =0 \label{eq15b} .
\end{align}
\end{subequations}
Equation (\ref{eq15b}) is a Robin boundary condition and valid only on $\partial M$. Note that this boundary condition has sufficient flexility by letting the parameter $m_s$ to be free. In fact, the term $(\zeta K +m_s)$ is a free parameter for the case of minimal coupling or $K$ being a constant on boundaries. This is the case for well-known boundary geometries like sphere, cylinder or planes. Putting $m_s\rightarrow\infty$ reinforces the Dirichlet B.C.. As we will see in the next section, for the case of parallel planes, and in fact for any configuration with constant $K$, it is possible to obtain the Neumann B.C. from (\ref{eq15b}) as well.

\subsection{The surface and volume energies}
Using (\ref{eq8a}) and \cite{Birrell}
\begin{equation}\label{eq30}
\langle 0|T_{\mu \nu }(x)|0\rangle =\sum_{\alpha }T_{\mu \nu }\{\phi {%
_{\alpha }(x),\phi _{\alpha }^{\ast }(x)\}},
\end{equation}
the vacuum expectation value $\langle0|T_{0}^{{\mathrm{(vol)}}0}|0\rangle$ has been obtained as follows \cite{Saharian2}
\begin{eqnarray} \label{eq31}
E^{{\mathrm{(vol)}}} &=&\int_{{\Sigma}}d^{D}x\sqrt{|g|}\langle
0|T_{0}^{{\mathrm{(vol)}}0}|0\rangle \nonumber \\
&=&\sum_{\alpha }\left\{ \frac{\omega _{\alpha }}{2}+\int_{\partial {\Sigma}%
}d^{D-1}x\sqrt{|g|}n^i\left[ (\zeta -1/4)\partial_i-\zeta (\partial _i \ln \sqrt{g_{00}})\right] \phi
_{\alpha }(x)\phi _{\alpha }^{\ast }(x)\right\} ,
\end{eqnarray}
in which $n^i$ is the spatial part of $n^\mu=(0,n^i)$, the unit normal to $\partial M_s$.

Same calculation for the surface energy will result in \cite{Saharian2}
\begin{eqnarray} \label{eq33}
E^{{\mathrm{(surf)}}} &=&\int_{{\Sigma}}d^{D}x\sqrt{|g|}\langle
0|T_{0}^{{\mathrm{(surf)}}0}|0\rangle \nonumber \\
&=&-\sum_{\alpha }\int_{\partial
{\Sigma}}d^{D-1}x\sqrt{|g|}n^i\left[ (\zeta -1/4)\partial
_i-\zeta (\partial _i\ln \sqrt{g_{00}})\right]
\phi _{\alpha }(x)\phi _{\alpha }^{\ast }(x).
\end{eqnarray}
Equation (\ref{eq31}) is quite important since the second term in (\ref{eq31}) is nothing but the surface energy $E^{{\mathrm{(surf)}}}$. In fact, we see that $E=E^{{\mathrm{(vol)}}}+E^{{\mathrm{(surf)}}}=\frac{1}{2}\sum_{\alpha}\omega _{\alpha }.$
As is shown in Appendix A (see after (\ref{eq23})), since $n^\mu\eta_\mu=0$ for any static spacetime, the total energy $E$ is a conserved quantity.

Calculation of $E^{{\mathrm{(surf)}}}$ or $E^{{\mathrm{(vol)}}}$ is not as easy task as one might suspect at first. For such a complicated and lengthy process see for example Ref.\cite{BorzooCQG} and references therein.

In the next subsection, we see that the surface energy vanishes under Dirichlet B.C.s in the spacetime described by (\ref{eq3}). We show the same statement for the case of parallel planes under Neumann B.C.s. This is an interesting results according to which it suffices simply to sum over mode frequencies $\omega_\alpha$ instead of complicated calculation of (\ref{eq31}) which is usually done in the literature.

\section{the vacuum energy for Neumann B.C.}
In this section we find the total energy $E$ for two conducting parallel planes subject to a class of gravitational fields described by metric (\ref{eq3}) which first used in \cite{BorzooEPJC}. It can be shown that any static, and even stationary, weak gravitational field can be transformed into the form (\ref{eq3}) by using perturbation expansion of the metric in the space between the planes\cite{Sorge}. The coordinate system is described by $(x,y,z,t)$ with the origin on one of the planes, i.e. the one nearest to the source of the gravitational field. The planes are separated by a distance $l$ in such a way that each plane is perpendicular to $z-$axis. Therefore, the planes are characterized by $z=0$ and $z=l$ respectively.

\subsection{The mode functions}
To find $\phi$, we notice the translational symmetry in $x$ and $y$ directions between the plates and use the ansatz
\begin{eqnarray}\label{eq39}
\phi(x) = C(\omega,k_\perp) e^{-i\omega t} e^{ik_x x}e^{ik_y y}Z(z),
\end{eqnarray}
for the wave function. Thus, the Klein-Gordon equation simplifies to
\begin{eqnarray}\label{eq40}
Z''(z)+\partial_z\ln{|\sqrt{-g}g^{11}|} Z'(z)+(-g^{00}g_{11}\omega^2-k_{\perp}^2+m^2g_{11})Z(z)=0,
\end{eqnarray}
for a typical diagonal static spacetime. Here $k_{\perp}^2\equiv k_x^2+k_y^2$. Note that the coupling to curvature has been ignored in the Klein-Gordon equation as it can be easily shown that the Riemann's scalar contributes only within second order perturbations. We use the solution to (\ref{eq40}) which has been found by Nazari \cite{BorzooCQG} as follows
\begin{eqnarray}\label{eq41}
Z(z)=C(\omega,k_\perp)\left(1-(\frac{\lambda}{2}+\frac{a}{4b})z\right)\sin\left(\sqrt{b}z(1+\frac{a}{4b}z)+\Theta_0 \right),
\end{eqnarray}
in which
\begin{subequations}
\begin{align}
a&=-2(\lambda_0-\lambda_1)\omega^2-2\lambda_1m^2, \label{eq42a}\\
b&=(1-2(\gamma_0-\gamma_1))\omega^2-k_\perp^2-(1+2\gamma_1)m^2,  \label{eq42b}\\
\lambda & = \lambda_1+\lambda_0. \label{eq42c}
\end{align}
\end{subequations}
Evidently, parameters $\omega$ and $\Theta_0$ should be determined by imposition of boundary conditions. After using the Neumann B.C. and doing some algebra it founds that
\begin{subequations}
\begin{align}
&\cot(\Theta_0)=\frac{1}{\sqrt{b}}(\frac{\lambda}{2}+\frac{a}{4b})=O(\lambda)=\epsilon, \label{eq43a}\\
&\cot(\sqrt{b}l+\frac{a}{4\sqrt{b}}l^2+\Theta_0)=\frac{1}{\sqrt{b}}\left(\frac{\lambda}{2}+\frac{a}{4b}\right)=\epsilon,  \label{eq43b}
\end{align}
\end{subequations}
which in turn gives
\begin{eqnarray}\label{eq44}
\sqrt{b}l+\frac{a}{4\sqrt{b}}l^2=n\pi,   \; \Theta_0=\frac{\pi}{2}-\epsilon .
\end{eqnarray}
Further algebra on this equations results in
\begin{subequations}
\begin{align}
\omega&=(1+\gamma_0-\gamma_1+\frac{\lambda_0-\lambda_1}{2}l)\sqrt{\omega_0^2+M^2} , \label{eq45a}\\
\Theta_0&=\frac{\pi}{2}-\epsilon, \label{eq45b}
\end{align}
\end{subequations}
in which $\omega_0^2=k_\perp^2+(\frac{n\pi}{l})^2$ and $M^2=(1+2\gamma_1+\lambda_1 l)m^2$. The same calculation shows that (\ref{eq45a}) holds also for Dirichlet condition.

\subsection{Energy for Dirichlet B.C.}
By substituting (\ref{eq39}) into (\ref{eq33}) we find
\begin{eqnarray} \label{eq33new}
E^{{\mathrm{(surf)}}} =-\sum_{\omega,k_\perp}C^2\int_{\partial
{\Sigma}}d^{D-1}x\sqrt{|g|}n^i\left[ (2\zeta -1/2)\partial
_i Z(z)-\zeta (\partial _i\ln \sqrt{g_{00}})Z(z)\right]
Z(z).
\end{eqnarray}
Since for Dirichlet B.C. we have $Z(z)|_{\partial M}=0$, quite generally the surface energy vanishes regardless of which material configuration and background (static) spacetime is used.

\subsection{Energy for Neumann B.C.}
The Neumann B.C., i.e. $n^\mu\partial_\mu \phi|_{\partial M}=0$, is compatible with (\ref{eq15b}) by setting either $\zeta=m_s=0$ or more generally $m_s=-\zeta K$. We take a short analysis on both cases.
\subsubsection{$\zeta=m_s=0$}
In this case, the surface energy momentum tensor (\ref{eq8b}) vanishes upon imposing the Neumann B.C.. Thus, for massless minimally coupled scalar field the total energy equals the zero point energy of the field, i.e. $E=\sum\hbar \omega/2$. This result is also a general statement regardless of which material configuration and background (static) spacetime is used.
\subsubsection{$m_s=-\zeta K$}
As $m_s$ and $\zeta$ are constants, this case is impossible unless $K$ be a constant which means that $\partial M_s$ should have constant extrinsic curvature. This is the case for some well-known geometries like parallel planes, sphere, cylinder and so on.
In this case, the first term containing $\partial_i Z(z)$ vanishes due to Neumann B.C. and we see from (\ref{eq33new}) that
\begin{eqnarray} \label{eq33neumann}
\begin{split}
E^{{\mathrm{(surf)}}} &=\zeta \sum_{\omega,k_\perp}C^2\int_{\partial \Sigma}d^{D-1}x\sqrt{|g|}(n^i\partial _i\ln \sqrt{g_{00}})Z^2(z) \\
&=\zeta \sum_{\omega,k_\perp}C^2\int_{upper \; plate}+\zeta \sum_{\omega,k_\perp}C^2\int_{lower \; plate}+\zeta \sum_{\omega,k_\perp}C^2\int_{side \; surf.},
\end{split}
\end{eqnarray}
where side surfaces are perpendicular to the $z-direction$. The normal vector $n^i$ is given by
\begin{eqnarray} \label{normalvector}
n^\mu|_{up.}=+\frac{\delta^\mu_3}{\sqrt{|g_{33}|}},\;\; n^\mu|_{low.}=-\frac{\delta^\mu_3}{\sqrt{|g_{33}|}},\;\
; n^\mu|_{x-dir.}=+\frac{\delta^\mu_1}{\sqrt{|g_{11}|}},\;\; n^\mu|_{y-dir.}=+\frac{\delta^\mu_2}{\sqrt{|g_{22}|}}.
\end{eqnarray}
By this, we mean that the vacuum state is that of an observer which is hovering in a spacetime point and static relative to the plates. Such an observer is inevitably accelerated relative to the source of the gravity. The energy on surfaces perpendicular to the $x$ and $y$ directions vanishes. In fact, in the $x$-direction, $n^i=(0,1,0,0)$ and we saw in (\ref{eq33}) that $n^i \partial_i \phi \phi^*=\partial_x \phi \phi^*=0$ because $\phi \phi^*$ is independent of $x$ according to (\ref{eq39}). The same argument is true for $y$-direction. Therefore, the third term on right side of (\ref{eq33neumann}) vanishes and the rest is simplified as follows:
\begin{eqnarray} \label{eq33neumann1}
E^{{\mathrm{(surf)}}} &=A\zeta\lambda_1 \sum_{\omega,k_\perp}C^2 \left\{g_{33}\sqrt{g_{00}}Z^2(z)\right\}|_{z=l}-A\zeta\lambda_1 \sum_{\omega,k_\perp}C^2 \left\{g_{33}\sqrt{g_{00}}Z^2(z)\right\}|_{z=0}.
\end{eqnarray}
By using (\ref{eq41}) we find for the first term that
\begin{eqnarray} \label{eq33neumann2}
\begin{split}
&A\zeta\lambda_1 \sum_{\omega,k_\perp}C^2 \left\{g_{33}\sqrt{g_{00}}Z^2(z)\right\}|_{z=l}\\
&=A\zeta\lambda_1 \sum_{\omega,k_\perp}C^2 \left(1-(\lambda+\frac{a}{2b})l\right)\sin^2\left(\sqrt{b}l+\frac{a}{4\sqrt{b}}l+\Theta_0 \right)\\
&=A\zeta\lambda_1 \sum_{\omega,k_\perp}C^2,
\end{split}
\end{eqnarray}
in which we have used (\ref{eq44}) and $\lambda_1 \left(1-(\lambda+\frac{a}{2b})l\right)=\lambda_1+O(\lambda^2)$. Note that $\sin^2(\Theta_0)=1+O(\epsilon^2)$. Some algebra shows that the second term in (\ref{eq33neumann1}) is also equal to (\ref{eq33neumann2}) hence $E^{{\mathrm{(surf)}}}=0$.

\subsubsection{Volume energy for Neumann and Dirichlet B.C.s}
So far we have shown that the total surface energy vanishes for both Neumann and Dirichlet boundary conditions and the conserved volume (total) energy reads
\begin{equation}\label{eq48}
E=\frac{1}{2}\sum_{\alpha}\omega _{\alpha }=(1+\gamma_0-\gamma_1+\frac{\lambda_0-\lambda_1}{2}l)\; \frac{1}{2}\int \frac{k_\perp dk_\perp}{(2\pi)^2} \sum_{n}\sqrt{\omega_0^2+M^2},
\end{equation}
in which we have used (\ref{eq45a}). The massive problem will be calculated in section (V). For now, we assume $M=0$. So, the summation on the right side of (\ref{eq48}) is nothing but the famous case of the Casimir effect in flat spacetime which can be regularized using standard procedures \cite{MostepanenkoBook} and the result is as follows
\begin{equation}\label{eq49}
E=-(1+\gamma_0-\gamma_1+\frac{\lambda_0-\lambda_1}{2}l)\; \frac{\pi^2 \hbar c}{1440 l^3},
\end{equation}
which in terms of proper distance
\begin{equation}\label{eq4950}
l_p=\int_0^l \sqrt{-g_{33}}dz=l(1+\gamma_1+\frac{1}{2}\lambda_1l),
\end{equation}
between the planes can be written as
\begin{equation}\label{eq50}
E=-(1+\gamma_0+2\gamma_1+\frac{\lambda_0+2\lambda_1}{2}l_p) \frac{\pi^2 \hbar c}{1440 l_p^3}.
\end{equation}
This result is in agreement with the previous studies which have used a different approach. See Eq.(82) in \cite{BorzooCQG} and the discussions after. Note that equation (\ref{eq50}) is valid also for $\zeta \neq 0, \; m \neq 0$ while it was obtained in \cite{BorzooCQG} for $\zeta=0, \; m=0$. Another consistency check is Eq. (5.4) in \cite{Esposito} which has been found using a special case of the metric (\ref{eq3}), i.e. with $\gamma_0=\gamma_1=\lambda_1=0,\; \lambda_0=g,$ and $ \; \zeta=0, \; m=0$.

Note that the first order constants $\gamma_0$ and $\gamma_1$ are present in (\ref{eq50}). We return to this point in the next section. Before that, we need to carry out some estimations.

\subsubsection{Some estimations }
The weak field limit of the the Schwarzschild metric in isotropic form is given by
\begin{eqnarray}\label{eq58}
ds^2 = (1 -\frac{R_s}{r}) dt^2- (1 +\frac{R_s}{r})d\Omega^2,
\end{eqnarray}
where $R_s=2\frac{GM}{c^2}$ is the Schwarzschild radius. Suppose the planes are located in a distance $R$ from the center of the source. If $0\leq z \leq l$ and the planes be small enough, we may expand \cite{Sorge}
\begin{eqnarray}\label{eq59}
\frac{1}{r}=\frac{1}{R+z}=\frac{1}{R}-\frac{1}{R^2}z+O(R^{-2}).
\end{eqnarray}
Thus, the metric (\ref{eq58}) recasts into
\begin{eqnarray}\label{eq60}
ds^2 = (1 +\frac{R_s}{R}-\frac{R_s}{R^2}z) dt^2- (1 -\frac{R_s}{R}+\frac{R_s}{R^2}z)d\Omega^2,
\end{eqnarray}
which in comparison with (\ref{eq3}) gives $\gamma_0=-\gamma_1=\frac{R_s}{2R}, \; \lambda_0=-\lambda_1=-\frac{R_s}{2R^2}$ and the energy reads
\begin{equation}\label{eq61}
E=-(1-\frac{R_s}{R}+\frac{R_s}{2R^2}l_p)\;\frac{\pi^2 \hbar c}{1440 l_p^3}.
\end{equation}
This shows that the energy increases and the leading order correction is of order $\frac{R_s}{R}\approx 7\times10^{-10}$ for the Earth while previous studies \cite{Sorge2019,Calloni,Sorge,Milton} predict the leading order correction to be $\frac{R_s l_p}{R^2}\approx 1.1\times10^{-22}$.

The force by which the planes attract each other is given by
\begin{eqnarray}\label{eq62}
F=-\frac{\partial E}{\partial l_p}=-(1+\gamma_0+2\gamma_1-\frac{\lambda_0+2\lambda_1}{3}l_p)\;\frac{\pi^2 \hbar c}{480 l_p^4},
\end{eqnarray}
which for the Schwarzschild spacetime gives
\begin{eqnarray}\label{eq63}
F=-\frac{\partial E}{\partial l_p}=-(1-\frac{R_s}{R}-\frac{R_s}{3R^2}l_p)\;\frac{\pi^2 \hbar c}{480 l_p^4}.
\end{eqnarray}
This shows that the mutual force between the planes decreases for Schwarzschild spacetime.\\

\section{more analysis on the presence of first order corrections $\gamma_0$ and $\gamma_1$ in the energy}
Although the energy in (\ref{eq50}) has been found in a quite unambiguous manner, one may still think that the constants $\gamma_0$ and $\gamma_1$ should not be present in the energy (\ref{eq50}) because we can initially absorb them into the time and space parts of the metric (\ref{eq3}) by using a simple rescaling $t\rightarrow (1+\gamma_0)t, \; \overrightarrow{x}\rightarrow (1+\gamma_1)\overrightarrow{x}$.
This idea has been used in some previous studies in the subject and we show that it is not correct. For instance see equation (8) in \cite{BorzooEPJC}, the discussion before equation (17) in \cite{Fulling} , equation (2.5) in \cite{Sorge2019} and equation (2.15) in \cite{Lima}. In the following, we give more elementary and satisfactory analysis to justify the presence of the constants $\gamma_0$ and $\gamma_1$ in (\ref{eq50}).

First we should observe that the above-mentioned rescaling is acceptable only when there is no boundary in the problem under consideration or the boundary be at spatial infinity. When the spacetime manifold has boundary, any transformation should be applied to the boundary as well, otherwise the physics will change. In fact, to achieve consistent results, any transformation must be applied to the whole problem (which consists of the metric, the boundaries and the Klein-Gordon equation) rather than only to the metric.
Second, to be more clear, we now apply the rescalings to both the spacetime and boundaries and calculate mode frequencies again.

To proceed, let's apply the transformations to the Klein-Gordon equation first. By $t^\prime=(1+\gamma_0)t, \; \overrightarrow{x}^\prime=(1+\gamma_1)\overrightarrow{x}$ we find
\begin{subequations}
\begin{align}
\partial_0&=(1+\gamma_0)\partial_{0^\prime} \;, \; \partial^2_0=(1+2\gamma_0)\partial^2_{0^\prime} \;, \label{eq51a}\\
\partial_i&=(1+\gamma_1)\partial_{i^\prime} \;, \; \partial^2_i=(1+2\gamma_1)\partial^2_{i^\prime} \;, \label{eq51b} \\
g_{00}&=(1+2\gamma_0)g_{0\prime 0^\prime} \;, \; g^{00}=(1-2\gamma_0)g^{\prime00} \; , \label{eq51a}\\
 g_{11}&=(1+2\gamma_1)g^\prime_{33}\;, \; g^{33}=(1-2\gamma_1)g^{\prime11} \; . \label{eq51b}
\end{align}
\end{subequations}
Now the metric recasts into
\begin{eqnarray}\label{eq52}
ds^2 = (1 +2 \lambda_0 z^\prime) dt^{\prime2}- (1+2 \lambda_1 z^\prime)\left(dx^{\prime2}+dy^{\prime2}+dz^{\prime2}\right),
\end{eqnarray}
It can be easily checked out that the Klein-Gordon equation is invariant under the rescalings. Thus we use
\begin{eqnarray}\label{eq53}
g^{\prime00}\partial_{t^\prime}^2\Phi+g^{\prime33}(\partial_{x^\prime}^2\Phi+\partial_{y^\prime}^2\Phi)
+\frac{1}{\sqrt{-g^\prime}}\partial_{z^\prime}(\sqrt{-g^\prime}g^{\prime33}\partial_{z^\prime}\Phi)+m^2\Phi=0.
\end{eqnarray}
The transformation changes the boundaries as well
\begin{eqnarray}\label{eq54}
\begin{split}
&z=0\;\rightarrow z^\prime=0, \\
&z=l\;\rightarrow z^\prime=l^\prime=l(1+\gamma_1).
\end{split}
\end{eqnarray}
For the wave function we have
\begin{eqnarray}\label{eq55}
\begin{split}
\Phi(x) &= C e^{-i\omega t} e^{i\textbf{k}_\perp.\textbf{x}_\perp}Z(z)= C e^{-i\omega(1-\gamma_0)(1+\gamma_0) t} e^{i\textbf{k}_\perp(1-\gamma_1).(1+\gamma_1)\textbf{x}_\perp}Z(z) \\
 &=C e^{-i\omega^\prime t^\prime} e^{i\textbf{k}_\perp^\prime . \textbf{x}^\prime_\perp}Z(z^\prime),
\end{split}
\end{eqnarray}
where
\begin{subequations}
\begin{align}
\omega^\prime&=\omega(1-\gamma_0), \label{eq56a}\\
k_\perp^\prime&=k_\perp(1-\gamma_1). \label{eq56b}
\end{align}
\end{subequations}
Now, the new problem consists of the metric (\ref{eq52}), the Klein-Gordon equation (\ref{eq53}) and the boundary condition (\ref{eq54}). To find mode frequencies we use the ansatz in (\ref{eq55}) and do the same process as already done through (\ref{eq39}) to (\ref{eq45a}). The result is given by
\begin{eqnarray}\label{eq57}
\omega^\prime&=(1+\frac{\lambda_0-\lambda_1}{2}l^\prime)\sqrt{\omega_0^{\prime2}+M^{\prime2}} ,
\end{eqnarray}
where $\omega_0^\prime=(k_\perp^{\prime2}+(\frac{n\pi}{l^\prime})^2)^{\frac{1}{2}}$ and $M^{\prime2}=(1+\lambda_1 l)m^2$. In (\ref{eq56a}) we saw $\omega=(1+\gamma_0)\omega^\prime$. Equations (\ref{eq54}) and (\ref{eq56b}) also give $\omega_0^{\prime2}=(1-2\gamma_1)(k_\perp^2+(\frac{n\pi}{l})^2)$. Moreover, note that $M^{\prime2}=(1-2\gamma_1)M^{2}$. Substituting all these back into (\ref{eq57}) recovers (\ref{eq45a}) again.

The non-vanishing corrections in first order perturbation is not related to the vacuum state. As we see in various spacetimes, for example, in Schwrazschild with $\gamma_0=-\gamma_1 \neq 0$ or in Kerr spacetime with $\gamma_0\neq -\gamma_1 \neq 0$, the corrections does not vanish. Since the vacuum state is different for this examples, we conclude that the non-vanishing contribution of $\gamma_0$ and $\gamma_1$ to the energy is not a vacuum effect. It is really a boundary effect and the main concern of this section is to justify such a point. In fact in the presence of boundary we do not eligible to perform gauge transformations to throw away and ignore the constants $\gamma_0$ and $\gamma_1$.


Another point is that the total energy (\ref{eq50}) does not depend on the coupling constant $\zeta$. Although we ignored $\zeta$ previously through the term $\zeta R$ , we expect to see it in the final result for the energy because $\zeta$ presents in the energy-momentum tensor. However, contrary to this expectation, we saw that $\zeta$ presents only in the surface sector of the energy-momentum tensor which was shown that vanishes for D.N. B.C.s.

\section{Energy for massive and finite temperature scalar field}
Regularizing and renormalizing the summation in equation (\ref{eq48}) follows standard procedures. However, it differs from that of usually done by $M=(1+\gamma_1+\frac{l}{2}\lambda_1 )m$ which contains not only $m$ but also the spacetime parameters $\gamma_1,\;\lambda_1$.  Therefore, we need to carefully redo the procedure to find the explicit contribution of $\gamma_1$ and $\lambda_1$. In fact, the energy appears in (\ref{eq50}) has the form  $(1+\gamma_0+2\gamma_1+...)E^{flat}_{m=0}$ in which $E^{flat}_{m=0}=-\frac{\pi^2}{1440 l_p^3}$ is the massless flat space Casimir energy. For the massive case, the Casimir energy has the same form with $E^{flat}_{m=0}$ being replaced by $E^{flat}_{m\neq0}$.

The summation in (\ref{eq48}) can be written as \cite{MostepanenkoBook}

\begin{eqnarray}\label{eq68}
\begin{split}
\mathbb{E}&=\frac{\hbar}{2}\int \frac{k_\perp dk_\perp}{(2\pi)^2} \sum_{n}\sqrt{\omega_0^2+M^2}\\
&=\frac{\hbar}{2}\int_0^\infty \frac{k_\perp dk_\perp}{2\pi} \left[ \sum_{n=0}^\infty \sqrt{k^2_\perp+\left(\frac{n\pi}{l}\right)^2+M^2}- \frac{l}{\pi}\int_0^\infty dk_z \sqrt{k^2_\perp+k_z^2+M^2}\right] \nonumber
\end{split}
\end{eqnarray}
\begin{eqnarray}\label{eq681}
=\frac{\pi^2}{4l^3}\int_a^\infty y dy  \left[ \sum_{n=0}^\infty \sqrt{y^2+n^{2}}- \int_0^\infty dt \sqrt{y^2+t^2}\right],
\end{eqnarray}
in which $a=\frac{l}{\pi}M,\; \hbar=c=1$. Using the Abel-Plana formulae
\begin{eqnarray}\label{eq69}
\sum_0^\infty F(n)-\int_0^\infty F(t)dt=\frac{1}{2}F(0)+i \int_0^\infty \frac{F(it)-F(-it)}{e^{2\pi t}-1}dt,
\end{eqnarray}
it can be shown that
\begin{eqnarray}\label{eq70}
\begin{split}
\mathbb{E}=-\frac{\pi^2}{6(2\pi)^4l^3} \int_{p_0}^\infty  \frac{\left(u^2-p_0^2\right)^{\frac{3}{2}}}{e^u-1}du,
\end{split}
\end{eqnarray}
in which $p_0=2lM$. Now, the important observation is that
\begin{eqnarray}\label{eq71}
p_0=2lM=2(1-\gamma_1-\frac{\lambda_1}{2}l_p)l_p (1+\gamma_1+\frac{\lambda_1}{2}l_p)m=2l_pm,
\end{eqnarray}
where we have used $l=(1-\gamma_1-\frac{\lambda_1}{2}l_p)l_p$ again. Equation (\ref{eq71}) proves that $\mathbb{E}$ is free of spacetime parameters and $\mathbb{E}=E^{flat}_{m\neq0}$ which is the massive Casimir energy in flat spacetime \cite{MostepanenkoBook}. Thus, we have shown that
\begin{eqnarray}\label{eq72}
E_{m\neq0}=(1+\gamma_0+2\gamma_1+\frac{\lambda_0+2\lambda_1}{2}l_p)E^{flat}_{m\neq0}.
\end{eqnarray}

The above formalism can be extended to the case of the finite temperature field using the Matsubara formalism \cite{Matsubara}. Some studies has been done previously in the literature \cite{borzoothermal1,borzoothermal2,SERNELIUS}. The point is that it suffices to find modified frequencies and simply use the Matsubara summation formula for the field in its excited states. See equations (23)-(28) in \cite{borzoothermal1}.

\section{A glance on corner (joint) terms}
If we have a discontinuity or abrupt change in the unit normal to $\partial M_s$ then the last term in (\ref{eq7}) should be kept. Since our setting consists of two parallel plates, there is a discontinuity in the normal vector on transition from one plate to another one at the edges of the plates.

The role of corner terms in the quasi-local approach to gravitational action has been explored in many studies. Joint terms may be on the intersection of $\Sigma_1$(or $\Sigma_2$) and $\partial M_s$ or completely on $\partial M_s$ itself. The later fits our problem as we know that the discontinuity in $n^\mu$ breaks the $2D$ surface $\partial \Sigma$ into parts. To find the probable contribution of the corresponding joint terms in the surface stress-tensor we use the following typical form of the joint terms in the gravitational action \cite{BrownLawYork}
\begin{eqnarray}\label{eq73}
\frac{1}{2}\int_{\partial \Sigma}d^{D-1}x\sqrt{|\sigma|}\phi^{2}(u^\mu\delta n^{\mu}+\overline{n}_\mu \delta u^\mu)
\end{eqnarray}
This form equals the third term in (\ref{eq7}) in the case of timelike joint. Now assume that $\partial \Sigma$ be broken into four parts corresponding to four edges of the apparatus. Each of this edges give $\delta n^{\mu}=n^{\mu}|_{z=l}-n^{\mu}|_{z=0}=n^{\mu}-(-n^{\mu})=2n^{\mu}$. On the other side, we know that $n_{\mu}u^\mu =0$ and consequently $u^\mu\delta n^{\mu}=0$. For the second term in (\ref{eq73}) we use $\delta u^\mu=0$. Therefore, the joint term contribution to the Casimir energy vanishes. The same manner can be employed for timelike joint terms in our setting \cite{Hayward,Lehner}.

\section{conclusion}
Since the Casimir effect is a phenomena which occur under the presence of boundaries, it is acceptable and natural to study it in curved spacetime through the quasi-local approach to the gravitational action and energy. In this paper, we applied the quasi-local approach to the Casimir energy and showed that the calculations become much easier and reliable. We found previous results in the literature (see after equation (\ref{eq50})) and extended the calculations to the case of massive and arbitrarily coupled scalar field. We found that for the case of massive field the energy equals the flat space Casimir energy times a factor composed of spacetime parameters (see section V). The coupling constant was absent in the final result for the energy due to the fact that it contributed only in the surface part of the energy. However, we showed that the surface energy vanishes under Dirichlet and Neuman boundary conditions. As an aside, the curvature coupling did not affect the energy up to second order perturbation under the influence of weak gravitational field. We showed that the above results are also valid for the case of Neumann boundary conditions.

The common thought was that the impact of the gravitational field on the Casimir energy appears only in second order perturbation expansion hence it can not be measured within the current precision of the experiments. We showed that the correction to the Casimir energy appears also within the first order perturbation calculations. As we shown in section IV, the source of error in previous studies was the fact that the lowest order perturbations had been omitted in the course of an unappropriate transformation of the spacetime metric. Therefore, the corrections are twelve orders of magnitudes larger than what previously found and hopefully be measured by the current precision of the experiments.

There was another issue related to the use of quasi-local approach for the Casimir effect. The joint terms should be incorporated in the gravitational action due to the presence of discontinuities in the normal vector of the boundaries. This terms may have contributions in the energy-momentum tensor and we shown that they vanish for the case of parallel plates.

\appendix
\section{conserved quantities}
If a spacetime admits some killing symmetries, one finds conserved quantities according to the Noether theorem. An important probable conserved quantity has been shown to be \cite{Saharian2}
\begin{equation}\label{eq22}
P_{\Sigma _2} - P_{\Sigma _1} = \int_{\partial M_s} d^D x
\sqrt{|h|} \, n_\mu n_\nu n^\lambda \eta _\lambda T^{{\rm (vol)}\mu\nu} ,
\end{equation}
where $P_{\Sigma }$ is defined by:
\begin{equation}\label{eq23}
P_{\Sigma } = \int_{{\Sigma}}d^{D}x\sqrt{|\gamma |} \,
u_\mu \eta _{\nu} T^{{\rm (vol)}\mu\nu}+ \int_{{\partial \Sigma}}
d^{D-1}x \sqrt{|\sigma |} \, \eta _{\nu} \overline{u}_{\mu} \tau ^{\mu\nu}.
\end{equation}
The first term in the right side is the volume part $P^{{\rm (vol)}}_{\Sigma }$ and the second one is $P^{{\rm (surf)}}_{\Sigma }$. The possibility to have $P_{\Sigma }$ conserved is available by letting $n^\lambda \eta _\lambda=0$ on the right side of (\ref{eq22}). This, in turn, needs to have some killing vectors tangent to boundaries of the spacetime manifold $M$. 

Any static spacetime has a timelike hypersurface-ortghogonal killing vector field. Suppose $\eta^\mu=\delta^\mu_0$ be such a killing vector. The hypersurface $\Sigma$ is described by $t=const.$ hence the unit normal vector is given by
\begin{equation}\label{eq28}
u^\mu=\frac{\delta^\mu_0}{\sqrt{|g_{00}|}},\; u_\mu=\sqrt{|g_{00}|}\,\delta_{\mu0},\; \eta_\mu=g_{\mu0},
\end{equation}
in which the hypersurface-orthogonal condition has been supposed to be $u^\mu=a\eta^\mu$ for some constant $a$. Putting (\ref{eq28}) into (\ref{eq23}) we find for the volume part of (\ref{eq23})
\begin{equation}\label{eq29}
\begin{split}
P^{{\rm (vol)}}_{\Sigma }&=\int_{{\Sigma}}d^{D}x\sqrt{|\gamma |} \,
u_\mu \eta _{\nu} T^{{\rm (vol)}\mu\nu}\\
&=\int_{{\Sigma}}d^{D}x\sqrt{|g|} \,T^{{\rm (vol)}0}_0,
\end{split}
\end{equation}
where use is made of $\sqrt{|g|}=\sqrt{|\gamma|}\sqrt{|g_{00}|}$ and $g^{00}=g_{00}^{-1}$ for a static spacetime. A similar result can be found for $P^{{\rm (surf)}}_{\Sigma }$.

\section *{Acknowledgments}
The author would like to thank University of Tehran for supporting this research under the grant No. 30102/1/01.

\end{document}